\documentclass[ a4paper,11pt]{revtex4}

\usepackage{multirow}
\usepackage{subfigure, amsfonts, amsmath}
\usepackage{longtable}
\usepackage{pdflscape}
\usepackage{float}
\usepackage{cancel}
\usepackage{graphicx}
\usepackage{array}
\usepackage{hyperref}
\usepackage{appendix}
\usepackage{amssymb}
\usepackage{epstopdf}
 \usepackage{multirow}
\usepackage{epsfig}
\usepackage{subfigure, amsfonts, amsmath}
\usepackage{longtable}
\usepackage{lscape}
\usepackage{subfigure}
\usepackage{amssymb}
\usepackage{amsfonts}
\usepackage{amsmath}
\usepackage{yhmath}
\usepackage[english]{babel}
\usepackage[latin1]{inputenc}
\usepackage[T1]{fontenc}
\usepackage{amsthm}
\usepackage{setspace}
\usepackage{tikz,pgflibraryshapes}

\newtheorem{theorem}{Theorem}

\newtheorem{axiom}{Axiom}

\newtheorem{corollary}[theorem]{Corollary}
\newtheorem{definition}[axiom]{Definition}

\newtheorem{proposition}{Proposition}

\renewcommand{\eqref}[1]{Eq.~(\ref{#1})}
\def\Xint#1{\mathchoice
   {\XXint\displaystyle\textstyle{#1}}%
   {\XXint\textstyle\scriptstyle{#1}}%
   {\XXint\scriptstyle\scriptscriptstyle{#1}}%
   {\XXint\scriptscriptstyle\scriptscriptstyle{#1}}%
   \!\int}
\def\XXint#1#2#3{{\setbox0=\hbox{$#1{#2#3}{\int}$}
     \vcenter{\hbox{$#2#3$}}\kern-.5\wd0}}

\def\dashint{\Xint-}

\begin{document}
\title{Dimensional Reduction without Continuous Extra Dimensions}

\author{Ali H. Chamseddine}
\affiliation{American University of Beirut, Physics Department, Beirut, Lebanon}
\affiliation{and I.H.E.S. F-91440 Bures-sur-Yvette, France}
\author{J. Fr\"ohlich, B. Schubnel}
\affiliation{ETHZ, Mathematics and Physics Departments, Z\"urich, Switzerland}
\author{D. Wyler}
\affiliation{Inst. of Theoretical Physics, University of  Z\"urich, Switzerland}
\begin{abstract}
We describe a novel approach to dimensional reduction in classical field theory. Inspired by ideas from noncommutative geometry, we introduce extended algebras of differential forms over space-time, generalized exterior derivatives and generalized connections associated with the "geometry" of space-times with discrete extra dimensions. We apply our formalism to theories of gauge- and gravitational fields and find natural geometrical origins for an axion- and a dilaton field, as well as a Higgs field.
\end{abstract}

\maketitle



\newpage

\section{Introduction}
Introducing extra dimensions in order to unify physical laws and identify natural geometrical origins of various gauge- and scalar fields has quite a long history, beginning in the 1920's with attempts by Kaluza and Klein (see [\onlinecite{KA}], [\onlinecite{KL}]) to unify Maxwell's theory with general relativity in a five-dimensional space-time, continuing with Pauli's construction of non-abelian SU(2)-gauge fields in a six dimensional space-time and culminating with string- and M-theory; (see, e.g., [\onlinecite{OR}]). All these attempts are plagued with the appearance of infinite towers of modes of ever larger mass. In theories where all modes are coupled to the gravitational field such towers may seem to be a problem.

Within the general framework of noncommutative geometry, Connes has proposed to consider generalized notions of differential geometry to describe extra dimensions and to construct classical field theories where certain scalar fields, such as the Higgs field of the standard model, appear for geometrical reasons, but towers of very massive modes do not arise; see [\onlinecite{CO1}], [\onlinecite{CO2}]. Connes' attempts are based on generalizations of spin geometry. The fundamental geometrical data are encoded in so-called "spectral triples", $(\mathcal{A}, D,\mathcal{H})$, where $\mathcal{A}$ is a (possibly non commutative) $^*$algebra of operators represented on a separable Hilbert space $\mathcal{H}$, and $D$ is an elliptic operator acting on $\mathcal{H}$ generalizing the Dirac operator.

In this note, we present an alternative approach to "dimensional reduction", based on certain extensions of the graded differential algebra,  $\Omega(M)$,  of differential forms over space-time M, that does not involve introducing continuous extra dimensions, but involves generalized notions of "exterior derivative", "connection" and "metric". Our approach is inspired by Connes' ideas ([\onlinecite{CO1}],[\onlinecite{CO2}]), but we attempt to generalize general Riemannian - rather than spin-geometry; (see [\onlinecite{FRO1}]). Thus, besides a *algebra of operators, it involves two anti-commutaing K\"ahler-Dirac operators, $\mathcal{D}$ and $\bar{\mathcal{D}}$, acting on a Hilbert space of generalized differential forms (rather than a single Dirac operator acting on a Hilbert space of generalized spinors). Classical fields are identified with elements of a (sub-)space of "zero modes"on which $\mathcal{D}^2=\bar{\mathcal{D}}^{2}$. The linear combinations $d:=\mathcal{D}-i \bar{\mathcal{D}}$ and $d^{*}:=\mathcal{D}+i\bar{\mathcal{D}}$ can then be interpreted as generalizations of the exterior derivative and its adjoint; (see [\onlinecite{FRO1}]).

The purpose of our note is to provide natural geometrical interpretations of various scalar fields, such as an axion-, a dilaton and a Higgs field, using ideas and results from [\onlinecite{FRO1}]. As in Connes' approach, "space-time" will have the structure of two copies of the usual four-dimensional space-time carrying ( a priori massless) left-handed and right-handed spinors, respectively. This is reminiscent of a five-dimensional generalization of the quantum Hall effect discussed in [\onlinecite{FRO2}], the extra fifth dimension being treated as a discrete two-point set.

The axion will turn out to be the "fifth" component of the electromagnetic vector potential, the dilaton to be a gravitational degree of freedom associated with the discrete fifth dimension, and the Higgs field will appear as a component of the electroweak gauge field that induces tunneling processes between the two sheets of "space-time"and  provides masses to the fermions and to the W- and Z gauge bosons, as sketched in figure 1. 

Our paper is organized as follows. In section II, we summarize, in a  sketchy  way, some elements of noncommutative geometry that are needed in subsequent sections. For further details, the reader is referred to [\onlinecite{CO1}],[\onlinecite{CO2}] and [\onlinecite{FRO1}].
In section III, we first recover an axion field (section \ref{axion}) by identifying it with the fifth component of the electromagnetic vector potential. This represents the simplest application of our formalism. In section \ref{dilaton}, we proceed to generalize the Einstein-Hilbert gravitational action to our two-sheeted space-time and find that this leads to the appearance of a dilaton field. Finally, in section \ref{Higgs}, we show how the Higgs field of the electroweak theory finds a natural geometrical interpretation within our formalism. Some additional remarks and conclusions are sketched in section IV. 

 \begin{figure}[H]
 \label{fig1}
\begin{center}
\begin{tikzpicture}[scale=0.8]
\draw[-,thick] (4,-2) -- (11,-7);
\draw[-,thick] (6,0) -- (13,-4.5);
 \draw (13,-4.5) node[below] {$\psi_R$};
 \draw(11,-7) node[below] {$\psi_L$};
 
\draw[<->] (7.5,-4.3) to[bend left=60] (9.7,-2.5);
 \draw(8.5,-3) node[below] {$\phi$};
  \end{tikzpicture}
   \caption{A schematic view of the Yukawa coupling between left- and right-handed fermions, interpreted in a five dimensional space time. The left- and right-handed fermions live on separate  four-dimensional sheets.  The Higgs field couples left- to right-handed spinors via quantum tunnelling.}
\end{center}
\end{figure}
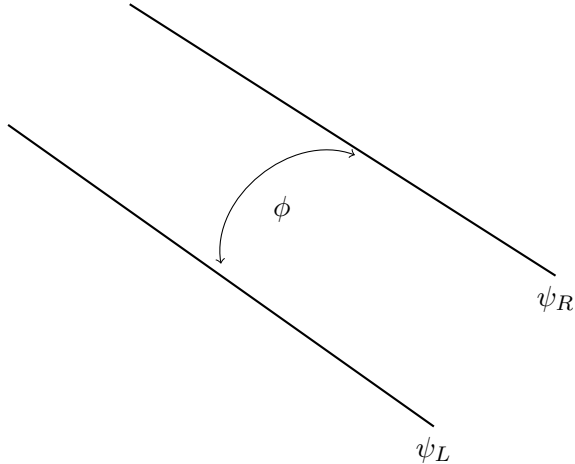

\section{Generalized differential geometry}

Gauge theories are intimately related to differential geometry. The reader may remember an undergraduate course on electromagnetism where the Maxwell equations were entirely rewritten in terms of differential forms.  Classical fields in a gauge theory with gauge group $G$ are sections of  some vector bundles over space-time $M$ associated to a  principal $G$-bundle   over $M$. Gauge potentials (such as  the U(1)- or SU(2)- gauge potentials) are $\mathcal{G}$-valued one-forms appearing in the definition of covariant derivatives in a  local basis of sections of associated vector bundles, and $\mathcal{G}$ is the Lie algebra of $G$.  A well-known theorem of Serre and Swan  ([\onlinecite{SW}]) tells us that all finite-dimensional  vector bundles  over a smooth compact manifold $M$  correspond to finitely generated projective $\mathcal{C}^{\infty}(M,\mathbb{C})$-modules. This result motivates our present approach. To generalize classical gauge theories, we  will  introduce $^{*}$algebras $\mathcal{A}$ (in particular non-commutative algebras) generalizing the commutative algebra $\mathcal{C}^{\infty}(M,\mathbb{C})$, and then  consider finitely generated projective $\mathcal{A}$-modules and define a generalization of the $\mathbb{Z}$-graded algebra of differential forms over $M$.  This furnishes the right kind of geometrical data enabling us to generalize the notion of gauge theories.

\subsection{Basic definitions}

Let $\mathcal{A}$ be  a unital $^{*}$algebra over the field $K=\mathbb{R}$ or $\mathbb{C}$. We denote by $\Omega(\mathcal{A})=\bigoplus_p \Omega^{p}(\mathcal{A})$ any $\mathbb{Z}$-graded differential algebra with $\mathcal{A} = \Omega^{0}(\mathcal{A}) $. The graded product over the "algebra of generalized differential forms" $\Omega(\mathcal{A})$ is  denoted  by $\omega\omega'$, where $\omega$, $\omega'$ are elements of $\Omega(\mathcal{A})$. The degree of a homogeneous element $\omega \in \Omega(\mathcal{A})$ is denoted by $\text{deg}(\omega)$. 

\begin{definition}
Vector bundles over $\mathcal{A}$
\end{definition}
Inspired by the theorem of Serre and Swan, one defines a noncommutative vector bundle,  $\mathcal{M}(\mathcal{A})$,   over $\mathcal{A}$ as a finitely generated  projective (left) $\mathcal{A} -$module (see [\onlinecite{CO1}]). Every such module   admits a generating family, i.e.,  there exist   $s_{1},...,s_{n} \in \text{Hom}(\mathcal{M}(\mathcal{A}),\mathcal{A})$, $e_{1},...,e_{n} \in \mathcal{M}(\mathcal{A})$ such that,  for all $x \in \mathcal{M}(\mathcal{A})$,
$$x=\sum_{i=1}^{n} s_{i}(x)e_{i}.$$ 

The set  $\lbrace e_{i} \in \mathcal{M}(\mathcal{A}) , \text{ }i=1,...,n \in \mathbb{N}\rbrace$,   is called a generating family  of  sections of the vector bundle $\mathcal{M}(\mathcal{A})$.
\bigskip

Next, we assume that there exists a $\mathbb{Z}_{2}$-graded nilpotent operator $d_{\mathcal{A}}$ ($d_{\mathcal{A}}^{2}=0$)  acting on  a $\mathbb{Z}$-graded   differential algebra $\Omega(\mathcal{A})$. Since $\Omega(\mathcal{A})$ is  a left $\Omega(\mathcal{A})$-module,
we may  define the  differential
\begin{equation}
\label{delta}
 \delta_{\mathcal{A}}:= \left[d_{\mathcal{A}}, \cdot  \right]_{g}
 \end{equation}
 
 on the algebra $\Omega(\mathcal{A})$, where the commutator $\left[ \cdot , \cdot \right]_{g}$ respects the $\mathbb{Z}_{2}$-grading of $\Omega(\mathcal{A}) $, i.e., 
 \begin{equation}
 \left[d_{\mathcal{A}},\omega_{p}\right]_{g}=d_{\mathcal{A}}\omega_{p}+(-1)^{p+1} \omega_{p}d_{\mathcal{A}}
 \end{equation}

 for any  $\omega_{p}$ of degree $p$.  For all homogeneous $ \omega \in \Omega(\mathcal{A})$,  we may assume that  $d_{\mathcal{A}} \omega$ is homogeneous.   Note that $\delta_{\mathcal{A}}$ is nilpotent:
\begin{eqnarray*}
\delta_{\mathcal{A}}^{2} \omega&=&d_{\mathcal{A}}^2 \omega+ (-1)^{\text{deg}(\omega)+1} d_{\mathcal{A}} \omega d_{\mathcal{A}}  + (-1)^{\text{deg}(d_{\mathcal{A}} \omega)+1} d_{\mathcal{A}} \omega d_{\mathcal{A}} +  \omega d_{\mathcal{A}}^2\\
&=&d_{\mathcal{A}}^2 \omega+ \omega d_{\mathcal{A}}^2=0.
\end{eqnarray*}

Furthermore, $\delta_{A}$  obeys the Leibniz's rule
\begin{equation}
\delta_{\mathcal{A}}( \omega \omega')=\delta_{\mathcal{A}}(\omega)  \omega'+(-1)^{\text{deg}(\omega)} \omega   \delta_{\mathcal{A}}\omega'
\end{equation}

and $\delta_{\mathcal{A}}(1_{\mathcal{A}})=0$. 
\bigskip

\begin{definition}
Connections
\end{definition}
Let $\mathcal{M}(\mathcal{A})$ be a projective, finitely generated (left) $\mathcal{A}$-module,  and let $\delta_{\mathcal{A}}$ be defined as in (\ref{delta}). A  connection, $\nabla$, on  $\mathcal{M}(\mathcal{A})$  associated to $\delta_{\mathcal{A}}$  is a $\mathbb{C}$-linear map
\begin{eqnarray*}
\nabla: \mathcal{M}(\mathcal{A}) &\longrightarrow&\Omega^{\text{odd}}(\mathcal{A})\otimes_{\mathcal{A}} \mathcal{M}(\mathcal{A})
\end{eqnarray*}

such that, for all  $a \in \mathcal{A}$, $s \in \mathcal{M}(\mathcal{A})$,
\begin{equation}
\label{Lei}
\nabla(as)=\delta_{\mathcal{A}} a \otimes s + a\nabla s.
\end{equation}

$\delta_{\mathcal{A}} a$ in (\ref{Lei}) is understood as $(\delta_{\mathcal{A}} a)1_{\mathcal{A}}=d_{\mathcal{A}} a -a d_{\mathcal{A}}1_{\mathcal{A}}$. 
Every projective  finitely generated module having a generating family $\lbrace e_{i} \rbrace_{i=1}^{n} $ of sections,  connections are entirely determined by their action on the $e_i$'s
$$\nabla(e_i)=- \Omega^{j}_{i} \otimes e_j,$$
where $\Omega^{j}_{i} \in \Omega^{\text{odd}}(\mathcal{A})$. The forms $\Omega^{j}_{i} $ correspond  to the gauge potential in classical gauge theories.
If the module is free and the generating family is a basis, one can choose arbitrary forms $ \Omega^{j}_{i} $. If the module is not free one has to impose some restrictions on the coefficients  $\Omega^{j}_{i}$ ([\onlinecite{CO1}],[\onlinecite{CH2}]).
\bigskip

We require that
\begin{equation}
\label{ca}
\nabla (\omega \otimes s)= \delta_{\mathcal{A}} \omega \otimes s + (-1)^{\text{deg}(\omega)} \omega  \nabla s
\end{equation}

for all homogeneous $\omega \in \Omega(\mathcal{A})$, $s \in \mathcal{M}(\mathcal{A})$, where the product  is between forms, i.e., $\omega (\omega_{1} \otimes s)= (\omega \omega_{1}) \otimes s$. As in (\ref{Lei}), $\delta_{\mathcal{A}} \omega$ in  (\ref{ca}) is understood as $(\delta_{\mathcal{A}} \omega)1_{\mathcal{A}}$.  Using (\ref{ca}), we can extend the definition of a connection to $\Omega(\mathcal{A}) \otimes_{\mathcal{A}}  \mathcal{M}(\mathcal{A})$ in a unique way and define  curvature as follows.

\begin{definition}
Curvature
\end{definition}
The curvature of a  connection $\nabla$ is the left $\mathcal{A}$-linear map:
\begin{equation}
\label{curv}
-\nabla^{2}:\mathcal{M}(\mathcal{A}) \longrightarrow \Omega^{\text{even}}(\mathcal{A})\otimes_{\mathcal{A}} \mathcal{M}(\mathcal{A}).
\end{equation}

\subsection{Generalization of  the algebra of differential forms}
Let $\mathcal{A}$ and   $\mathcal{B}$ be  unital algebras over the  field $K=\mathbb{R}$ or $\mathbb{C}$. We consider $\mathbb{Z}$-graded differential algebras $\Omega(\mathcal{A})$ and $\Omega(\mathcal{B})$,  with $\mathcal{A}=\Omega^{0}(\mathcal{A})$, $\mathcal{B} = \Omega^{0}(\mathcal{B})$. We write $\mathcal{C}:=\mathcal{A} \otimes_{K} \mathcal{B}$. Then $\Omega(\mathcal{A}) \otimes_{K} \Omega(\mathcal{B})$ is a left $\mathcal{C}$-module and can be equipped with a graded product. Henceforth we usually omit the  "$K$" in $\otimes_{K}$.

\bigskip

\begin{definition}
Graded product over $\Omega(\mathcal{A}) \otimes \Omega(\mathcal{B})$
\end{definition}
The graded product, $ \wedge$, over the algebra $\Omega(\mathcal{A}) \otimes \Omega(\mathcal{B}) $ is defined as follows:  For all homogeneous elements $\omega,\omega' \in \Omega(\mathcal{A}) $ and $ \sigma,\sigma' \in \Omega(\mathcal{B}) $,
\begin{equation}
(\omega \otimes \sigma) \wedge (\omega' \otimes \sigma') = (-1)^{\text{deg}(\sigma) \text{deg}(\omega')}  \omega \omega ' \otimes \sigma  \sigma'.
\end{equation}

With this product, $\Omega(\mathcal{A}) \otimes \Omega(\mathcal{B})$ is a $\mathbb{Z}$-graded algebra, and we have that
$$(\Omega(\mathcal{A}) \otimes \Omega(\mathcal{B}))^{n}=\underset{p+q=n}{\bigoplus} \Omega(\mathcal{A})^{p}  \otimes \Omega(\mathcal{B})^{q} $$

where $\Omega^{p}(.)$ is the subspace of $\Omega(.)$ of degree $p$.
\bigskip

We  assume that there exist $\mathbb{Z}_{2}$-graded  nilpotent operators $d_{\mathcal{A}}$ on $\Omega(\mathcal{A})$ and $d_{\mathcal{B}}$ on $\Omega(\mathcal{B})$.
\bigskip

\begin{definition}
Extension of $(d_{\mathcal{A}}, d_{\mathcal{B}})$
\end{definition}
An extension of $(d_{\mathcal{A}}, d_{\mathcal{B}})$ is a  $\mathbb{Z}_{2}$-graded, linear nilpotent operator  $\tilde{d}$ acting on  the  left $\mathcal{C}$-module $\Omega(\mathcal{C}):=\Omega(\mathcal{A}) \otimes \Omega(\mathcal{B})$ that can be written in the form
\begin{equation}
\label{dA}
\tilde{d} = \alpha d_{\mathcal{A}} \otimes 1_{\mathcal{B}} + \beta  \Gamma_{\mathcal{A}} \otimes d_{\mathcal{B}} + \sigma
\end{equation}

where $\sigma$ is an odd element of  $\Omega(\mathcal{C})$, $\alpha, \beta \in K$, and $\Gamma_{\mathcal{A}}$ is the involution on $\Omega(\mathcal{A})$ defined by
 $$ \Gamma_{\mathcal{A}}( \omega)=(-1)^{\text{deg}(\omega)} \omega, $$

 for a homogeneous $\omega \in \Omega(\mathcal{A})$. 
\bigskip

 As in (\ref{delta}), we  define a differential $\tilde{\delta}:=\left[\tilde{d}, \cdot \right]_{g}$ on the graded algebra $\Omega(\mathcal{C})$, as well as  connections and  curvature on any (noncommutative) vector bundle $\mathcal{M}(\mathcal{C})$. When $\sigma=0$ in (\ref{dA}), it is easy to check that $\tilde{d}^{2}=0$. Let $\kappa:=\omega \otimes \omega' \in \Omega(\mathcal{C})$, with $\omega$ homogeneous. One then has that 
\begin{eqnarray*}
\tilde{d}^2\kappa&=&\tilde{d} ( \alpha d_{\mathcal{A}} \omega \otimes \omega' + \beta (-1)^{\text{deg}(\omega)} \omega \otimes d_{\mathcal{B}} \omega')\\
&=& \alpha^2 d_{\mathcal{A}}^{2}\omega \otimes \omega' + \alpha \beta (-1)^{\text{deg}(d_{\mathcal{A}} \omega)} d_A  \omega \otimes d_{\mathcal{B}} \omega' + \alpha \beta (-1)^{\text{deg}(\omega)} d_{\mathcal{A}} \omega \otimes d_B \omega' +  \beta^2 \omega \otimes d_{\mathcal{B}}^{2}\omega'\\
&=&0.
\end{eqnarray*}

If $\sigma \neq 0$ one must add the conditions that $\left[\alpha d_{\mathcal{A}} \otimes 1_{\mathcal{B}} + \beta  \Gamma_{\mathcal{A}} \otimes d_{\mathcal{B}} , \sigma \right]_{g}=0$ and $\sigma^{2}=0$.
\bigskip

Below, we will choose for $\Omega(\mathcal{B})$ the exterior algebra of a finite-dimensional vector space $V$ over $K$, which we denote by $\mathcal{G}(V)$; ( $\mathcal{G}$ stands for "Grassmann Algebra"). This is a graded commutative algebra over the field $K$.  The algebra $\mathcal{B}$ is the field $K$. We denote by $\times $ the exterior product on $\mathcal{G}(V)$, and, with $\mathcal{C}=\mathcal{A} \otimes_{K} K \approx \mathcal{A}$, $\Omega(\mathcal{A})_{V}:=\Omega(\mathcal{C})=\Omega(\mathcal{A}) \otimes \mathcal{G}(V)$. 
\bigskip

Let $\xi_p \in \mathcal{G}(V)$ be a homogeneous element of odd degree $p$. The operator $d_{\mathcal{B}}:=\xi_p \times(.)$ acting on $\mathcal{G}(V)$ is linear, $\mathbb{Z}_{2}$-graded and  nilpotent. For  any $\alpha, \beta \in K$,
\begin{equation}
\label{da2}
\tilde{d}:=\alpha d_{\mathcal{A}} \otimes 1 + \beta  \Gamma_{\mathcal{A}} \otimes (\xi_p \times .)=\alpha d_{\mathcal{A}} \otimes 1 + \beta  (1_{\mathcal{A}} \otimes \xi_p) \wedge \cdot 
\end{equation}

is linear, $\mathbb{Z}_{2}$-graded and  nilpotent. More generally, we have the following proposition.
\bigskip

\begin{proposition}
Let $\xi_p \in \mathcal{G}(V)$ be a homogeneous element of odd degree,  and, let $\omega \in \Omega(\mathcal{A})$ be an even differential form such that $\delta_{\mathcal{A}}(\omega)=0$.  Then,  for all $\alpha \in K$,
\begin{equation}
\label{da3}
\tilde{d}= \alpha d_{\mathcal{A}} \otimes 1+ (\omega \otimes \xi_p) \wedge \cdot
\end{equation}

is a linear  nilpotent $\mathbb{Z}_{2}$-graded operator on $\Omega(\mathcal{A})_{V}$.
\end{proposition}

  If  $\omega \in Z(\Omega(\mathcal{A}))$ (the center of  $\Omega(\mathcal{A})$) and if the vector space $V$ is one dimensional, then  $\tilde{\delta}=\left[d_{\mathcal{A}} \otimes 1 + (\omega \otimes \xi_1) \wedge, .\right]_{g}$  maps $\Omega(\mathcal{A}) \otimes 1$ to itself, 
$$\tilde{\delta} (\omega' \otimes  1)=\delta_{\mathcal{A}} \omega' \otimes  1$$
for any $\omega' \in \Omega(\mathcal{A}) $.  The action of  $\tilde{\delta}$  on $\Omega(\mathcal{A}) \otimes \xi_1$  is also of the form $\tilde{\delta} (\omega' \otimes  \xi_1)=\delta_{\mathcal{A}} \omega' \otimes  \xi_1$. In other words, $\tilde{\delta}=\delta_{\mathcal{A}} \otimes  1$.
\bigskip

\begin{corollary}
Let $\omega_{i}$, $i=1,...,n$, be commuting differential forms  of homogeneous even  degree in $\Omega(\mathcal{A})$ such  that  $\delta(\omega_i)=0$, and let $\xi_{p_i} \in \mathcal{G}(V)$ be homogeneous elements of odd degree. Then,  for all $\alpha \in K$,
\begin{equation}
\tilde{d}= \alpha d_{\mathcal{A}}  \otimes 1 + \sum_{i} (\omega_i \otimes \xi_{p_{i}}) \wedge \cdot
\end{equation}

is a linear  nilpotent $\mathbb{Z}_{2}$-graded operator on $\Omega(\mathcal{A})_{V}$.
\end{corollary}

\subsection{Hermitian structure on $\Omega(\mathcal{A})_{V}$}
\label{sec3C}
Until now, the algebras $\mathcal{A}$   and  $\Omega(\mathcal{A})$  have been quite general. From now on, we focus on the case where $\mathcal{A}=\mathcal{C}^{\infty}(M ,M_{n}(K))$ and $\Omega(\mathcal{A})=\Omega(M ,M_{n}(K))$ is the $\mathbb{Z}$-graded algebra of $M_{n}(K)$- ($n \times n$ matrices) valued forms, where $M$ is a compact, orientable, smooth manifold of dimension $m$.  In this section, we suppose that $M$ is Riemannian. The exterior derivative  on $\Omega(M ,M_{n}(K))$ is denoted by $d$.
We construct a hermitian structure  on  $\Omega^{p}(\mathcal{A})_{V}$ using a generalization of  the Hodge operator on $\Omega(\mathcal{M})$. 
Let $\text{dim}(V)=k$. We choose a basis $(\xi^{m+1},...,\xi^{m+k})$ of $V$. This basis has the same properties as the fermionic superspace coordinates used in the theory of supermanifolds. We  introduce the notion of  Berezin integration on $\mathcal{G}(V)$  well known from fermionic functional integrals.
\bigskip

\begin{definition}
Berezin integration
\end{definition}
Let $\int_{b}$ denote  Berezin integration on $\mathcal{G}(V)$, i.e.,
$\int_{b}  d\xi^{i} \xi^{i}=1 $, $\int_{b}   d\xi^{i} =0 $, and
\begin{equation}
\int_{b} d\xi^{m+k} ... \text{ } d\xi^{m+1}  \xi^{m+1} ... \text{ } \xi^{m+k}= 1.
\end{equation}

Take $(dx^{1},...,dx^{m})$ to be  a coordinate basis of $1$-forms on $M$. To define the extended  Hodge operator, we write: $\xi^{i}:=dx^{i} \otimes 1 \equiv dx^{i}$. The metric on the manifold M is denoted by $g$. To raise the   indices of the totally  antisymmetric tensor $\epsilon_{\mu \nu ...}$, we extend $g$ by imposing $g^{(m+i)j}=g^{j(m+i)}=\delta^{(m+i) j}$, for all $i \in \lbrace{1,...,k\rbrace}$ and $j \in \lbrace{1,...,m+k\rbrace}$. This choice is consistent because  it is not affected by any change of coordinates on $M$.
\bigskip

\begin{definition}
Extended Hodge operator
\end{definition}
The extended Hodge $*$-operator is the  map $*.:\Omega^{p}(\mathcal{A})_{V} \longrightarrow \Omega^{m+k-p}(\mathcal{A})_{V}$ defined by:
\begin{equation}
\label{Ho}
*(\xi^{\mu_{1}} \wedge...\wedge \xi^{\mu_{p}})= \frac{\sqrt{\mid g \mid}}{(m+k-p)!} \epsilon^{\mu_{1}...\mu_{p}}_{\text{   } \text{   }\nu_{p+1}...\nu_{m+k}} \xi^{\nu_{p+1}} \wedge...\wedge \xi^{\nu_{m+k}} 
\end{equation}

and if $\omega=\frac{1}{p!} \omega_{\mu_{1}...\mu_{p}} \xi^{\mu_{1}} \wedge...\wedge \xi^{\mu_{p}} \in \Omega^{p}(\mathcal{A})_{V} $, by
\begin{equation}
\label{Ho2}
*(\omega)=\frac{\sqrt{\mid g \mid}}{(m+k-p)! p!} (\omega_{\mu_{1}...\mu_{p}})^{\dagger} \text{   } \epsilon^{\mu_{1}...\mu_{p}}_{\text{   } \text{   }\nu_{p+1}...\nu_{m+k}} \xi^{\nu_{p+1}} \wedge...\wedge \xi^{\nu_{m+k}},
\end{equation}

where $^{\dagger}$ is the adjoint on $M_{n}(K)$. 
\bigskip

Next, we  construct a hermitian structure  $(\cdot ,\cdot )$ on the  $\mathcal{A}$-module $\Omega^{p}(\mathcal{A})_{V}$, for any $p \in \mathbb{N}$. A hermitian structure is a sesquilinear form  $(\cdot ,\cdot ):\Omega^{p}(\mathcal{A})_{V} \times \Omega^{p}(\mathcal{A})_{V} \rightarrow \mathcal{A}$, such that
\begin{equation}
\label{Her}
    \begin{array}{ll}
         i)& (as,bs')=a(s,s')b^{\dagger} \mbox{,  for all }  a,b \in \mathcal{A},\text{  } s,s' \in \Omega^{p}(\mathcal{A})_{V}, \\
        ii)& (s,s) \ge 0, \mbox{ for all } s \in \Omega^{p}(\mathcal{A})_{V} \mbox{,  and } (s,s)=0 \Rightarrow s=0. \\ 
    \end{array}
\end{equation}

For arbitrary $\omega, \omega' \in \Omega^{p}(\mathcal{A})_{V}$, we define $( \cdot , \cdot )$ by
\begin{equation}
\label{her2}
\omega \wedge (* \omega')=:(\omega, \omega') d\mathcal{V}
\end{equation}

where $d \mathcal{V}= \sqrt{\mid g \mid} \xi^{1}... \text{ }\xi^{m+k}$ is the invariant extended volume form. The fact that  $( \cdot , \cdot )$  defined in  (\ref{her2}) satisfies properties i) and ii) of (\ref{Her}) is obvious from the definitions.
\bigskip
\bigskip

The space $\Omega^{p}(\mathcal{A})_{V}$ of $p-$forms also carries a scalar product
\begin{eqnarray*}
\langle \cdot ,  \cdot \rangle: \Omega^{p}(\mathcal{A})_{V} \times \Omega^{p}(\mathcal{A})_{V} &\longrightarrow&  \mathbb{C} \cup \lbrace \pm \infty \rbrace \\
(\omega, \omega') &\longmapsto&   \langle \omega ,  \omega' \rangle:=\dashint tr ( \omega \wedge *(\omega')),  
\end{eqnarray*}

where we have set 
\begin{equation}
\dashint \omega:=(-1)^{mk} \int_{M} \int_{b} d\xi^{m+k}... \text{ } d\xi^{m+1}  \omega.
\end{equation}

The factor $(-1)^{mk}$ ensures positivity of the scalar product and comes from the anticommutation relations $\lbrace \xi^{i}, \xi^{j}\rbrace=0$,  for $i,j \in \lbrace 1,...,m+k \rbrace$. On the right-hand side, the Berezin integration is defined in the following way: For all $ \omega=\frac{1}{p!}\omega_{\mu_{1}...\mu_{p} } \xi^{\mu_{1}}...\text{ } \xi^{\mu_{p}} \in \Omega^{p}(\mathcal{A})_{V}$,
\begin{equation}
 \int_{b} d\xi^{m+k}...  \text{ } d\xi^{m+1}   \omega  :=\frac{1}{p!} \omega_{\mu_{1}...\mu_{p} } \int_{b} d\xi^{m+k}... \text{ } d\xi^{m+1} \xi^{\mu_{1}}... \text{ }\xi^{\mu_{p}}
 \end{equation}
 
and the Berezin integration is carried out by putting all the Berezin variables $\xi^{\mu_{i}}$ on the left after passing them through the coordinate 1-forms. For instance, $$\int_{b} d\xi^{1}  dx dy \xi^{1}= \left(\int_{b} d\xi^{1} \xi^{1} \right) dx dy=dx dy.$$

 \section{Dimensional Reduction}
 
In this section, we apply "generalized differential geometry"  to some examples from classical field theory in order to show that various  classical  fields, such as the axion,  acquire a natural geometrical interpretation. We begin with the axion field that has appeared in [\onlinecite{FRO2}]  by dimensional reduction of Maxwell theory, starting from a five-dimensional bulk space-time.

\subsection{Axion field}
\label{axion}
To recover the axion field, only  a little change of  the differential geometric formulation of electromagnetism is necessary. Let $M$ be a compact four-dimensional Lorentzian manifold without boundary.  We consider the algebra $\mathcal{A}=\mathcal{C}^{\infty}(M, \mathbb{C})$  ($K=\mathbb{C}$). The new ingredient that makes  the axion field appear is the modification of the graded algebra of differential forms over $M$. We choose $V=\lbrace \lambda \xi_1, \lambda \in \mathbb{C} \rbrace $ the one-dimensional vector space spanned by $\xi_1$, and its exterior algebra $\mathcal{G}(V)$. On $\Omega(\mathcal{A})_{V}=\Omega(M,\mathbb{C}) \otimes\mathcal{G}(V)$, we define a natural generalization of the exterior derivative satisfying the hypotheses of Proposition 1. 
\begin{equation}
\tilde{d}=d \otimes 1 +\alpha (1 \otimes \xi_1) \wedge \cdot
\end{equation}

with $\alpha \in \mathbb{C}$.
A connection $\nabla$  on  $\mathcal{M}(\mathcal{A}):=\mathcal{A}=\mathcal{C}^{\infty}(M, \mathbb{C})$ is a $\mathbb{C}$-linear map 
\begin{equation*}
\nabla : \mathcal{C}^{\infty}(M, \mathbb{C}) \longrightarrow \Omega^{1}(\mathcal{A})_{V} \otimes_{\mathcal{C}^{\infty}(M)} \mathcal{C}^{\infty}(M, \mathbb{C})  \cong \Omega^{1}(\mathcal{A})_{V}.
\end{equation*}
\bigskip

\begin{proposition}
Let $\nabla$ be any  connection on $\mathcal{C}^{\infty}(M, \mathbb{C})$ and $ f \in \mathcal{C}^{\infty}(M, \mathbb{C})$.   Then 
 \begin{equation}
 \nabla f= - \Omega \otimes f
 \end{equation}
 
  where $\Omega= \omega + \phi \xi_1$, with $ \omega \in \Omega^{1}(M, \mathbb{C})$, $\phi \in \mathcal{C}^{\infty}(M, \mathbb{C})$.
\end{proposition}
\bigskip

This proposition follows directly from the definition of $\nabla$. The module being free, we require that  $\phi \in \mathcal{C}^{\infty}(M, \mathbb{R})$,  so that the field $\phi$ has zero charge. $\phi$  will turn out to be the axion field. The curvature (see (\ref{curv})) associated to a  connection is
\begin{eqnarray*}
- \nabla^{2} f&=&- \nabla (- \Omega \otimes f) = (\tilde{\delta} \Omega) \otimes f  - \Omega \wedge \nabla f\\
&=&(\tilde{\delta} \Omega ) \otimes f =: F_{\nabla^{2}} \otimes f
\end{eqnarray*}
with 
\begin{equation}
F_{\nabla^{2}}=\tilde{\delta} \Omega =\left[\tilde{d},\omega + \phi \xi_1 \right]_{+} 1= d \omega + d \phi \xi_1. 
\end{equation}

In [\onlinecite{FRO2}], the integral of the Chern-Simons five-form led to an axion term in the action. The corresponding extended integral of the extended Chern-Simons five-form $\Omega \wedge F_{\nabla^{2}} \wedge F_{\nabla^{2}}$ is given by
\begin{eqnarray*}
\dashint \Omega \wedge F_{\nabla^{2}} \wedge F_{\nabla^{2}}&=&
\dashint (\omega + \phi \xi_1) \wedge \left( d \omega + d \phi  \xi_1 \right) \wedge  \left( d \omega + d \phi  \xi_1   \right)\\
&=&\dashint (\omega + \phi \xi_1) \left( (d \omega)^{2} + 2 d \omega  d \phi  \xi_1\right) \\
&=& \dashint\left( \omega d\omega d\omega + \phi \xi_1 (d \omega)^{2} + 2 \omega d \omega d \phi \xi_1 \right).
\end{eqnarray*}

The Berezin  integration $\int_b d \xi_1 =0$  implies that $ \dashint \omega d\omega d\omega=0$. 
\begin{eqnarray*}
\dashint \Omega \wedge F_{\nabla^{2}} \wedge F_{\nabla^{2}}&=& \dashint  \left( \phi  (d \omega)^{2} \xi_1 + 2 \omega d \omega d \phi \xi_1  \right)\\
&=&\int_{M} \left( \phi  (d \omega)^{2}  + 2 \omega d \omega d \phi  \right).
\end{eqnarray*}

The manifold $M$ has no boundary, and therefore 
$$0=\int_{M} d(\omega  d \omega \phi)=\int_{M} d \omega  d \omega  \phi - \int_M \omega d \omega d \phi,  $$
which finally yields
\begin{equation}
\label{axa}
\dashint \Omega \wedge F_{\nabla^{2}} \wedge F_{\nabla^{2}}=3 \int_{M}  \phi (d \omega)^{2}
 \end{equation}
 
with $d \omega$ in (\ref{axa})   the electromagnetic field strength in four-dimensional space-time.  We see that $\phi$ can be interpreted as an axion field that couples to the electromagnetic field. We find the same result as in  [\onlinecite{FRO2}]. However, we have not added any extra continuous dimension. We  recover the kinetic term for the axion by dimensional reduction of the Maxwell action
\begin{eqnarray*}
\dashint  F_{\nabla^{2}} \wedge  (*F_{\nabla^{2}})&=&\int_{M} d \omega \wedge *(d \omega)_{4}+
\dashint (d \phi  \xi_1) \wedge *(d \phi  \xi_1)\\
&=&\int_{M} d \omega \wedge *(d \omega)_{4}+\int_{M} \partial^{\mu} \phi \text{ } \partial_{\mu} \phi \sqrt{\mid g\mid}\text{ } d^4x
\end{eqnarray*}
where $*(.)_{4}$ is the Hodge operator on $\Omega(M)$. 

\subsection{Gravity with dilaton}
\label{dilaton}
We  derive an  Einstein-Hilbert action with dilaton using our formalism. We  consider a four-dimensional compact Lorentzian manifold $M$ without boundary  and choose $K=\mathbb{R}$, $\mathcal{A}=\mathcal{C}^{\infty}(M, \mathbb{R})$ and  $V=\lbrace \lambda \xi_1, \lambda \in \mathbb{R} \rbrace $. On $\Omega(\mathcal{A})_{V}:=\Omega(M,\mathbb{R}) \otimes\mathcal{G}(V)$, we take
\begin{equation}
\tilde{d}=d \otimes 1 +\alpha (1 \otimes \xi_1) \wedge \cdot.
\end{equation}

We consider the vector bundle $\mathcal{M}(\mathcal{A})=\Omega^{1}(\mathcal{A})_{V}$. It  generalizes the cotangent  bundle of the manifold  $M$. Connections, $\nabla$, on $\mathcal{M}(\mathcal{A})$ are linear maps:
  $$\nabla: \Omega^{1}(\mathcal{A})_{V} \longrightarrow \Omega^{1}(\mathcal{A})_{V} \otimes_{\mathcal{A}} \Omega^{1}(\mathcal{A})_{V}.$$

To keep  our notation simple  in the following calculations, we identify $\xi_1 \equiv dx^4 $, as if $\xi_1$ were the  coordinate one-form corresponding to an extra dimension.
 We introduce an  extension of the Cartan basis 
 \begin{equation}
 \label{cart}
 E^{A}=e^{A}_{C} dx^{C},
 \end{equation}
 
where $A,C=0,...,4$ and $ ( E^{A}, E^{B} )= \eta^{AB}$;  $( \cdot, \cdot)$ is the hermitian structure on  $\Omega^{1}(\mathcal{A})_{V}$ defined in  (\ref{her2}),  and $\eta^{AB}$ is the Minkowski  metric tensor in five dimensions with  signature $(-,+,+,+,+)$.
\bigskip

 \begin{proposition}
 Let $\nabla$ be a  connection on $\Omega^{1}(\mathcal{A})_{V}$.
With respect to the Cartan basis,  
\begin{equation}
\nabla E^{A}= - \Omega_{B}^{A} \otimes E^{B}
\end{equation}

 where $\Omega_{B}^{A} \in \Omega^{1}(M, \mathbb{R})$,  for $A,B \in \lbrace{0,...,4 \rbrace}$, i.e.,
\begin{equation}
 \Omega_{B}^{A}=\omega_{B}^{A} + \phi_{B}^{A} dx^{4}
 \end{equation}
 
  with $\omega_{B}^{A} \in \Omega^{1}(\mathcal{A})_{V}$, $\phi_{B}^{A} \in \mathcal{C}^{\infty}(M, \mathbb{R})$.
\end{proposition}
\bigskip

The curvature two form associated to $\nabla$ takes the form:
\begin{eqnarray*}
-\nabla^{2} (\alpha_{A} E^{A})&=&- \nabla((\tilde{\delta} \alpha_{A}) \otimes E^{A} - \alpha_{A} \Omega_{B}^{A} \otimes E^{B}) \\
&=&- \left[(\tilde{\delta} \alpha_{A} \wedge \Omega_{B}^{A}) \otimes E^{B}  - \tilde{\delta}(\alpha_{A} \Omega_{B}^{A} ) \otimes E^{B} -\alpha_{A} \Omega_{B}^{A} \wedge \Omega_{C}^{B} \otimes E^{C} \right]\\
&=& \alpha_{A} (\tilde{\delta} \Omega_{C}^{A} +\Omega_{B}^{A} \wedge \Omega_{C}^{B}) \otimes E^{C}= \alpha_{A} \mathcal{R}_{C}^{A} \otimes E^{C}
\end{eqnarray*}

where 
\begin{eqnarray}
\label{R}
\mathcal{R}_{C}^{A}&=&\tilde{\delta} \Omega_{C}^{A} + \Omega_{B}^{A} \wedge \Omega_{C}^{B}. 
\end{eqnarray}

We can compute the scalar curvature using (\ref{R}). In the following calculations, we denote by capital letters $A,B,...$ indices that take values in $\lbrace 0,1,2,3,4 \rbrace$ and by $a,b,...$ indices in the range $0$ to $3$. To simplify matters,  we suppose that the Cartan basis is of the form
\begin{equation}
E^{A}= \delta^{A}_{a} e^{a}_{\mu} dx^{\mu} + \delta^{A}_{4} e^{\sigma} dx^{4}
\end{equation}

where $\sigma \in \mathcal{C}^{\infty}(M,\mathbb{R})$. With this ansatz, we tacitly assume that the added dimension does not "warp" when one moves along  $M$. The hermitian structure  $( \cdot, \cdot)$  defined in (\ref{her2})  satisfies $(dx^{\mu}, dx^{4})=0$. 
We make the following hypotheses:
\begin{itemize}
\item The connection is torsion free, i.e.,  $T(\nabla)=0$; (for the definition of  torsion see [\onlinecite{FRO1}])
\item The connection is unitary with respect to the metric on the extended tangent space, i.e., 
\begin{equation}
\label{unit}
\tilde{\delta} (\omega_{1},\omega_{2})=(\nabla \omega_{1}, \omega_{2})+ (\omega_{1}, \nabla \omega_{2})
\end{equation}
 
for arbitrary $\omega_{1},\omega_{2} \in \Omega^{1}(\mathcal{A})_{V}$.
\end{itemize}

These constraints characterize  the Levi-Civita connection.
\subsubsection{Torsion-free condition}
One has that
\begin{eqnarray*}
T(\nabla) E^{A}&=& \tilde{\delta} E^{A} + \Omega^{A}_{B} \wedge E^B =0.
\end{eqnarray*}

By writing $e^{a}=e^{a}_{\mu} dx^{\mu}$, it is easy to show that this condition leads to the following identities.
\begin{itemize}
\item $A=a$: 
\begin{equation}
\label{tf1}
\left\{
    \begin{array}{ll}
        de^{a} + \omega_{ b}^{a} e^{b} &= 0 \\
        \omega_{\nu 4}^{a} e^{\sigma}  -  \phi_{b}^{a}   e^{b}_{\nu} &=0
    \end{array}
\right.
\end{equation}
\item $A=4$: 
\begin{equation}
\label{tf2}
\left\{
    \begin{array}{ll}
        \omega^{4}_{ \mu b } e^{b}_{\nu} - \omega^{4}_{ \nu b } e^{b}_{\mu}&= 0 \\
		(\partial_{\nu} \sigma ) e^{\sigma} + \omega_{\nu 4}^{4} e^{\sigma}  -  \phi_{b}^{4} e^{b}_{\nu}&=0
    \end{array}
\right.
\end{equation}
\end{itemize}

\subsubsection{Unitarity condition}
Next, we use the unitarity condition (\ref{unit})
\begin{eqnarray*}
\tilde{\delta} (E^A,E^B)&=&(\nabla E^{A}, E^B)+ (E^{A}, \nabla E^B)\\
&=& - (\Omega^{A}_{C} E^{C}, E^B) - (E^{A}, \Omega^{B}_{D} E^D)\\
&=&- \Omega^{A}_{C} \eta^{CB} -  \Omega^{B}_{D} \eta^{AD}.\\
\end{eqnarray*}

By definition, $$\tilde{\delta} (E^A,E^B) = \tilde{\delta} (  \eta^{AB} )=0.$$

Consequently, we are led to
\begin{equation}
\label{u1}
\left\{
    \begin{array}{ll}
        \omega^{A}_{C} \eta^{CB} + \omega^{B}_{D} \eta^{AD}&= 0 \\
		\phi^{A}_{C} \eta^{CB} + \eta^{AD} \phi^{B}_{D}	&=0
    \end{array}
\right.
\end{equation}

Listing  all the possibilities for the components $A$ and $B$ of $\omega$ and $\phi$, and using equations (\ref{tf1}) and (\ref{tf2}), we see that  the components of the connection satisfy the identities
\begin{equation}
\label{coef}
\left\{
    \begin{array}{ll}
		\phi^{4}_{b}&=e^{\nu}_{b} (\partial_{\nu} \sigma) e^{\sigma}\\
        \phi^{b}_{4}&= - e^{\nu}_{c} (\partial_{\nu} \sigma) e^{\sigma} \eta^{cb} 
    \end{array}
\right.
\end{equation}
with all other components of $\phi^{A}_{B}$ vanishing. For $\omega^{A}_{B}$, only the forms $\omega^{a}_{b}$ may be non zero.

\subsubsection{Components of the curvature tensor }
We  have to find an expression for the components of the curvature tensor in terms of the components of the connection calculated in (\ref{coef}).  According to (\ref{R}),
\begin{eqnarray*}
\mathcal{R}^{A}_{B}&=& \tilde{\delta} \Omega^{A}_{B} + \Omega^{A}_{C} \Omega^{C}_{B}\\
&=&\frac{1}{2} R^{A}_{BCD} \text{ } E^{C} \wedge E^{D}.
\end{eqnarray*}

An easy identification leads  to
\begin{eqnarray}
\label{curvv}
R^{A}_{Bcd }&=& e^{\mu}_{c} e^{\nu}_{d}  \left( \partial_{\mu} \omega_{\nu B}^{A} - \partial_{\nu} \omega_{\mu B}^{A}  + \omega_{\mu E}^{A} \omega_{\nu B}^{E} - \omega_{\nu E}^{A} \omega_{\mu B}^{E} \right)\\
R^{A}_{B4d}&=& e^{- \sigma} e^{\nu}_{d} \left(  - \partial_{\nu}\phi_{B}^{A}  + \phi_{E}^{A} \omega_{\nu B}^{E} - \omega_{\nu E}^{A} \phi_{B}^{E} \right).
\end{eqnarray}

As our main goal is to compute the scalar curvature, we have to find the components of the Ricci tensor  using that
\begin{equation}
\label{R2}
R_{BD}=R^{A}_{BAD}= R^{a}_{BaD} + R^{4}_{B4D}.
\end{equation}

Because the scalar curvature is given by 
\begin{equation}
\label{R3}
R= \eta^{BD} R_{BD},
\end{equation}

 we only have to determine $R_{bd}$ and $R_{44}$. 
For instance,
\begin{eqnarray*}
R_{bd}&=&R^{a}_{bad} + R^{4}_{b4d}\\
 &=& \underbrace{e^{\mu}_{a} e^{\nu}_{d}  \left( \partial_{\mu} \omega_{\nu b}^{a} - \partial_{\nu} \omega_{\mu b}^{a}  + \omega_{\mu c}^{a} \omega_{\nu b}^{c} - \omega_{\nu c}^{a} \omega_{\mu b}^{c} \right)}_{ = R^{(4)}_{bd}} +\\
 &+&  \underbrace{e^{- \sigma} e^{\nu}_{d} \left(  - \partial_{\nu}\phi_{b}^{4}  + \phi_{c}^{4} \omega_{\nu b}^{c} -\cancel{ \omega_{\nu c}^{4} \phi_{b}^{c}}  +  \cancel{\phi_{4}^{4} \omega_{\nu b}^{4}} - \cancel{\omega_{\nu 4}^{4} \phi_{b}^{4}}\right)}_{(I)}.\\
\end{eqnarray*}

It is possible to evaluate (I)  using  properties of the Cartan basis. One  finds that 
\begin{eqnarray*}
(I)&=&   e^{\nu}_{d}  \left( - \underbrace{\left[ \partial_{\nu} e^{\mu}_{b} -    \omega_{\nu b}^{c} e^{\mu}_{c} \right] }_{(II)} \partial_{\mu} \sigma  -  e^{\mu}_{b} (\partial_{\nu} \partial_{\mu} \sigma) -   e^{\mu}_{b} (\partial_{\nu} \sigma)( \partial_{\mu} \sigma)   \right).
\end{eqnarray*}

The term (II) underlined above reduces to $(II)=- e_{b}^{\alpha} \Gamma^{\mu}_{\nu \alpha}$  where $\Gamma^{\mu }_{ \nu \alpha} $ are the Christoffel symbols, defined,  in any coordinate basis,  by
$$\nabla^{(4)} (dx^{\mu})=-\Gamma^{\mu }_{ \nu \alpha}  dx^{\nu} \otimes  dx^{\alpha} $$
and  $\nabla^{(4)}$ is the Levi-Civita connection on $\Omega^{1}(M,\mathbb{R})$,  given by
$$\nabla^{(4)} (E^{a})=-\omega^{a}_{b}\otimes E^{b}.$$

Indeed,
\begin{eqnarray*}
\nabla^{(4)}(E^{a})&=&\nabla^{(4)}(e^{a}_{\kappa} dx^{\kappa})=\partial_{\nu} e^{a}_{\kappa} dx^{\nu} \otimes dx^{\kappa} - e^{a}_{\delta} \Gamma^{\delta }_{ \nu \kappa}  dx^{\nu} \otimes  dx^{\kappa}\\
&=&- \omega^{a}_{ \nu c} e^{c}_{\kappa} dx^{\nu} \otimes  dx^{\kappa}
\end{eqnarray*}

which yields
\begin{equation}
\label{8}
\partial_{\nu} e^{a}_{\kappa} - e^{a}_{\delta} \Gamma^{\delta }_{ \nu \kappa} = -\omega^{a}_{ \nu c} e^{c}_{\kappa}.
\end{equation}

Moreover, as $e^{a}_{\mu} e^{\mu}_{b}=\delta^{a}_{b}$, 
\begin{equation}
\label{9}
\partial_{\nu}e^{\mu}_{b}=-e^{\kappa}_{b} e^{\mu}_{a} \partial_{\nu} e^{a}_{\kappa}.
\end{equation}

Plugging (\ref{8}) and (\ref{9}) into (II), 
\begin{eqnarray*}
(II)&=&-e^{\kappa}_{b} e^{\mu}_{a} (-\omega^{a}_{ \nu c} e^{c}_{\kappa}+ e^{a}_{\delta} \Gamma^{\delta }_{ \nu \kappa} )- \omega_{\nu b}^{c} e^{\mu}_{c}\\
&=& - e^{\alpha}_{b} \Gamma^{\mu }_{ \nu \alpha}.
\end{eqnarray*}

Thus,  
\begin{eqnarray*}
(I)&=& -  e^{\nu}_{d}  e_{b}^{\mu} \left(  \underbrace{\left[  -\Gamma^{\alpha}_{\nu \mu} \partial_{\alpha} \sigma  +   \partial_{\mu} \partial_{\nu} \sigma \right]}_{= \nabla^{(4)}_{\mu} \partial_{\nu} \sigma} +    (\partial_{\nu} \sigma)( \partial_{\mu} \sigma)   \right)
\end{eqnarray*}

where we have identified the components of the covariant derivative of $\partial_{\nu} \sigma$. Then,  
\begin{equation}
R_{bd}= R_{bd}^{(4)}   -  e^{\nu}_{d}  e_{b}^{\mu} \left( \nabla^{(4)}_{\mu} \partial_{\nu} \sigma +    (\partial_{\nu} \sigma) ( \partial_{\mu} \sigma)   \right).
\end{equation}

In the same way, one finds for $R_{44}$
\begin{eqnarray}
R_{44}&=&   - e^{\nu}_{a}  e^{\mu}_{c}  \eta^{ca} \left[ \nabla^{(4)}_{\nu} \partial_{\mu} \sigma + (\partial_{\nu} \sigma) (\partial_{\mu} \sigma) \right].
\end{eqnarray}

Using (\ref{R3}), the extended scalar curvature is given by
\begin{equation}
R^{(5)}=R^{(4)} -  2 g^{\mu \nu} \left[ \nabla^{(4)}_{\nu} \partial_{\mu} \sigma + (\partial_{\nu} \sigma) (\partial_{\mu} \sigma) \right].
\end{equation}

\subsubsection{Einstein-Hilbert action and Dilaton}

The generalized  Einstein-Hilbert action reads
\begin{eqnarray*}
\dashint  \sqrt{\mid g \mid} \text{ }  R^{(5)} e^{\sigma} d^4x \xi_1 &=&\int_{M} d^{4} x \sqrt{\mid g \mid}   R^{(5)} e^{\sigma} =\int_{M} d^{4} x \sqrt{\mid g \mid}  e^{\sigma} \left( R^{(4)} -  2 g^{\mu \nu} \left[ \nabla_{\nu} \partial_{\mu} \sigma + (\partial_{\nu} \sigma) (\partial_{\mu} \sigma) \right]\right)
\end{eqnarray*}

where we have replaced $\nabla^{(4)}_{\nu}$ by $\nabla_{\nu}$,  as there is no risk of confusion,  anymore. One can use  a conformal transformation to change the form of the integrand. Suppose that we rescale the metric,
$$\tilde{g}_{\mu \nu} = e^{2 \Phi} g_{\mu \nu}.$$

For a manifold $M$ of dimension $d$ (cf. [\onlinecite{CHO}]), this rescaling changes the scalar curvature by 
$$e^{2 \Phi} \tilde{R} - R =-2(d-1) \nabla^{\nu} \partial_{\nu} \Phi - (d-2)(d-1)(\partial^{\nu} \Phi) (\partial_{\nu} \Phi). $$

 $M$  is four-dimensional and if we choose $\Phi=\frac{1}{2} \sigma$, we find that
$$e^{  \sigma} \tilde{R} - R =-3 \nabla^{\nu} \partial_{\nu} \sigma - \frac{3}{2}(\partial^{\nu} \sigma) (\partial_{\nu} \sigma).$$

Here $R=R^{(4)}$.  Consequently, the generalized  Hilbert-Einstein action is given by
\begin{eqnarray}
\int_{M} d^4 x \sqrt{\mid g \mid} e^{\sigma}\text{ }  R^{(5)}&=&\int_{M} d^{4} x \sqrt{\mid \tilde{g} \mid} \text{ }   \left( \tilde{R}  - \frac{1}{2} \tilde{g}^{\mu \nu}  (\partial_{\nu} \sigma ) (\partial_{\mu} \sigma )  + \tilde{\nabla}^{\nu} \partial_{\nu} \sigma \right).
\end{eqnarray}

$ \tilde{\nabla}^{\nu} \partial_{\nu} \sigma$ can be rewritten as $\frac{1}{ \sqrt{\mid \tilde{g} \mid}} \partial_{\nu} (  \sqrt{\mid \tilde{g} \mid}  \partial^{\nu} \sigma)$. As $M$ is without boundary, 
$$ \int_{M} d^{4} x \sqrt{\mid \tilde{g} \mid} \text{ }  \tilde{\nabla}^{\nu} \partial_{\nu} \sigma=0$$

 and only the kinetic term for the dilaton remains (cf. for instance [\onlinecite{DA}]):
\begin{equation}
S=\int_{M} d^{4} x \sqrt{\mid \tilde{g} \mid} \text{ }   \left( \tilde{R} - \frac{1}{2} \tilde{g}^{\mu \nu}  (\partial_{\nu} \sigma ) (\partial_{\mu} \sigma )\right).
\end{equation}
\bigskip

\subsection{ Electroweak theory with a  Higgs field}
\label{Higgs}
 Let M be a four-dimensional compact Lorentzian manifold without boundary and  $\mathcal{A}=\mathcal{C}^{\infty}(M,\mathbb{C})$. We consider the $\mathcal{A}$-bimodule $\tilde{\mathcal{M}}(\mathcal{A})=S^{2}(M)\otimes_{\mathcal{A}} \mathcal{M}(\mathcal{A})$ where $\mathcal{M}(\mathcal{A})=\mathcal{C}^{\infty}(M,\mathbb{C}^2 \oplus  \mathbb{C})$ and $S^{2}(M)$ is the Hilbert space of square integrable spinors on $M$. $\tilde{\mathcal{M}}(\mathcal{A})$  is projective and finitely generated. We consider the  one-dimensional vector space $V=\lbrace \lambda \xi_1  , \lambda  \in \mathbb{C}\rbrace$ and introduce the exterior derivative 
 \begin{equation}
 \tilde{d}= d\otimes 1 
 \end{equation}

 on $\Omega(\mathcal{A})_{V}$. Connections  on $\mathcal{M}(\mathcal{A})$ are linear maps
\begin{eqnarray*}
\nabla: \mathcal{M}(\mathcal{A}) &\longrightarrow&\Omega^{\text{odd}}(\mathcal{A})_{V}\otimes_{\mathcal{A}} \mathcal{M}(\mathcal{A}).
\end{eqnarray*}

Once we have constructed a connection on $\mathcal{M}(\mathcal{A})$, we can construct a connection on $\tilde{\mathcal{M}}(\mathcal{A})$ in the following way.  Let $\nabla_{S^2}$ be the canonical spin connection on $S^{2}(M)$. We define 
\begin{eqnarray*}
\tilde{\nabla}: \tilde{\mathcal{M}}(\mathcal{A}) &\longrightarrow&\Omega^{\text{odd}}(\mathcal{A})_{V}\otimes_{\mathcal{A}} \tilde{\mathcal{M}}(\mathcal{A})
\end{eqnarray*}

by
$$\tilde{\nabla}(\psi \otimes f)=\nabla_{S^2} \psi \otimes f + \pi(\psi \otimes \nabla f )$$

where $\pi(\psi \otimes \omega \otimes f)=\omega \otimes \psi \otimes f$,  for all $\psi \in S^{2}(M)$, $\omega \in \Omega(\mathcal{A})_{V}$ and $f \in \mathcal{M}(\mathcal{A})$. 
\bigskip

We  construct a connection $\nabla$ on the free $\mathcal{A}$-module $\mathcal{M}(\mathcal{A})$.  Let $(s_1,s_2, s_3)$  be a basis of $\mathcal{M}(\mathcal{A})$.
$$\nabla s_i = - \underbrace{\Omega^{j}_{i}}_{\in  \Omega^{1}(\mathcal{A}) _{V} } \otimes s_j$$

Similarly to (\ref{R}), the components of the curvature tensor are given by
\begin{equation}
(F_{\Omega})^{i}_{j}=\tilde{\delta} \Omega^{i}_{j}+ \Omega^{i}_{k} \Omega^{k}_{j}.
\end{equation}

The general form of $\Omega:=(\Omega^{j}_{i})$ reads, in matrix notation,
\begin{equation}
\label{om}
\Omega= A \otimes 1 + B \otimes \xi_1
\end{equation}

where $A \in \Omega(M,M_{3}(\mathbb{C})), B \in M_{3}(\mathbb{C})$. The module being free, we can take an arbitrary consistent choice for $A,B$ in (\ref{om}).  We first introduce a Hermitian structure on $\mathcal{M}(\mathcal{A})$ in which the basis is orthonormal, i.e., we choose $(\cdot,\cdot)$ such that $(s_{i},s_{j})=\delta_{ij}$ and $(f,f')=\sum_{i=1}^{3} f_{i} \bar{f}'_{i}$, $\bar{(\cdot)}$ denoting complex conjugation. We require $\Omega$ to be unitary with respect to this metric, i.e., $\Omega$ must be skew-hermitian. We would like $A$ to be chosen  as  in the Standard Model of particle physics (see, e.g. [\onlinecite{WE}]); i.e.,
$$A=  \left( \begin{array}{cc}
\omega_{ \tiny 2 \times 2}&0_{  2 \times 1}\\
0_{\tiny 1 \times 2} & \alpha_{  1 \times1}\\
\end{array}
\right)$$

where $\omega=\omega_{\mu}  dx^{\mu}$ and $\alpha_{1 \times1}=\alpha_{\mu} dx^{\mu}$ are  the  $U(2)$ and  $U(1)$ gauge potentials, respectively. The form $\Omega$ being skew-hermitian, $\omega_{\mu}$ must be skew-hermitian and  $\alpha_{\mu} \in i\mathbb{R}$.
We would like $B \notin M_{2}(\mathbb{C}) \oplus \mathbb{C}$ to  exchange left- and right-handed spinors, describing tunneling processes between the two sheets of space-time as explained in section I. 
\begin{eqnarray*}
B=  \left( \begin{array}{cc}
0_{ \tiny 2 \times 2}&H\\
-H^{\dagger}& 0_{  1 \times1}\\
\end{array}
\right)
\end{eqnarray*} 

where $H \in \mathcal{C}^{\infty}(M, M_{2 \times 1}(\mathbb{C}))$. 
We can  add to  $B$ an axion field $\phi \in \mathcal{C}^{\infty}(M,\mathbb{R})$, as considered in section \ref{axion}. Then the final form for $\Omega$ is given by 
\begin{equation}
\Omega=  \left( \begin{array}{cc}
\omega_{  2 \times 2}&0_{  2 \times 1}\\
0_{ 1 \times 2} & \alpha_{  1 \times1}\\
\end{array}
\right)  \otimes 1 + \left( \begin{array}{cc}
i \phi 1_{  2 \times 2}&H\\
-H^{\dagger}& i \phi\\
\end{array}
\right) \otimes \xi_1.
\end{equation}

Next, we determine the components of the curvature two-form.
Before doing so, we propose to  investigate how the components of $\Omega$ transform under a gauge transformation.

\subsubsection{Gauge transformations}
Consider two bases of sections, $\lbrace s'_i \rbrace $, $\lbrace s_j \rbrace $ such that 
$$s'_i=g^{j}_{i}s_j.$$

One has that
\begin{equation}
\label{gauge}
(\Omega')_{i}^{l}=- (\tilde{d} g^{k}_{i}) (g^{-1})^{l}_{k} + g^{j}_{i} \Omega^{k}_{j} (g^{-1})^{l}_{k}.
\end{equation}

The matrix-valued function $g$  maps a basis of sections to another basis of sections. Since eft- and right-handed spinors should not be mixed by gauge transformations,  the most general form for $g$ is 
$$g= \left( \begin{array}{cc}
A_{  2 \times 2}&  0\\
0& e^{i \theta}\\
\end{array}
\right), $$
where   $A \in U(2)$, $\theta \in \mathbb{R}$.
One then finds that
\begin{eqnarray*}
 \omega &\rightarrow & A \omega A^{\dagger} - dA A^{\dagger} \\
 \alpha  &\rightarrow& \alpha - i d \theta\\
 H &\rightarrow& A H e^{-i \theta}  \\
 \phi &\rightarrow& \phi.  \\
 \end{eqnarray*}
 
 $H$  transforms as the standard Higgs field under a gauge transformation.
 We will need these formulas to check that the gauge field strength transforms correctly under  gauge transformations, i.e.,
\begin{equation*}
F_{\Omega} \rightarrow g F_{\Omega} g^{-1}.
\end{equation*}

\subsubsection{Curvature 2-form}
We  use the  notations
\begin{eqnarray*}
D H&:=&d H + \omega H - \alpha H\\
F_{\omega}&:=&d \omega + \omega^2\\
F_{\alpha}&:=&d \alpha.
\end{eqnarray*}

Then
\begin{equation}
\label{FF2}
F_{\Omega}=
\left( \begin{array}{cc}
F_{\omega} +i (d \phi) \xi_{1} 1_{2 \times 2} & (D H) \xi_1 \\
-(D H)^{\dagger} \xi_1 &i d \phi \xi_{1} + F_{\alpha}
 \end{array}
\right).
\end{equation}

Under a gauge transformation $g$,  $F_{\Omega}$  given in (\ref{FF2}) satisfies the transformation law
 \begin{equation}
 \label{ga}
 F_{\Omega} \rightarrow \left( \begin{array}{cc}
A_{ 2 \times 2}&  0\\
0& e^{i \theta}\\
\end{array}
\right)  F_{\Omega}\left( \begin{array}{cc}
A^{\dagger}_{ 2 \times 2}&  0\\
0& e^{-i \theta}\\
\end{array}
\right).
\end{equation}

Indeed,
\begin{eqnarray*}
D H &\rightarrow& d (A H e^{-i \theta})   + (A \omega A^{\dagger} - dA A^{\dagger})(A H e^{-i \theta} ) - (\alpha - i d \theta )( A H e^{-i \theta})\\
&=& \cancel{d A H e^{-i \theta}} + e^{-i \theta} A dH   - \bcancel{i A e^{-i \theta} d\theta} + A \omega H e^{-i \theta} - \cancel{dA H e^{-i \theta}} - \alpha A H e^{-i \theta} + \bcancel{i (d \theta) A H e^{-i \theta}}\\
&=& A D H e^{-i \theta}.
\end{eqnarray*}

All the other components are easily determined. It follows from (\ref{ga}) that the action functional
\begin{equation}
\label{Ya}
S= \dashint tr \left[ F_{\Omega} \wedge (*F_{\Omega})\right]
\end{equation}
is gauge-invariant.

\subsubsection{Yukawa coupling and kinetic energy term for the Higgs field}

It is not difficult to compute  the Hodge dual of $F_{\Omega}$ in the  basis ($dx^0,dx^1,dx^2,dx^{3},dx^4=\xi_1)$. 
If $\omega=\frac{1}{p!} \omega_{\mu_{1}...\mu_{p}} dx^{\mu_{1}} \wedge...\wedge dx^{\mu_{p}} \in \Omega^{p}(M, M_3(\mathbb{C}))_{V} $, 
$$*(w)=\frac{\sqrt{\mid g \mid}}{(m+1-p)! p!} (\omega_{\mu_{1}...\mu_{p}})^{\dagger} \text{   } \epsilon^{\mu_{1}...\mu_{p}}_{\text{   } \text{   }\nu_{p+1}...\nu_{m+k}} dx^{\nu_{p+1}} \wedge...\wedge dx^{\nu_{m+k}}.$$

As in \ref{sec3C}, to raise the lower indices of $\epsilon_{A B ...}$, we extend the metric tensor $g$ of the manifold M by defining $g^{4 A}=\delta_{4}^{A}$, for $A \in \lbrace 0,1,2,3,4 \rbrace$. In what follows, $A, B,...$ are indices that range from 0 to 4,  whereas $\mu, \nu,...$ take values in $\lbrace 0,1,2,3 \rbrace$. We work in the  signature $(-,+,+,+)$ for the Minkowski metric $\eta_{\mu \nu}$.

If the two-form $\omega = \frac{1}{2} \omega_{\mu \nu} dx^{\mu} dx^{ \nu}$  does not  contain any term with $\xi_1 $ then
\begin{eqnarray*}
*(\omega)&=&\frac{\sqrt{\mid g \mid}}{3! 2!} (\omega_{\mu \nu})^{\dagger} \text{   } \epsilon^{\mu\nu}_{\text{   } \text{   }A...C} dx^{A} \wedge...\wedge dx^{C}\\
&=& \frac{\sqrt{\mid g \mid}}{2! 2!} (\omega_{\mu \nu})^{\dagger} \text{   } \epsilon^{\mu \nu}_{\text{   } \text{   }\delta \gamma} dx^{\delta} \wedge dx^{ \gamma} \wedge \xi_1\\
&=& *(\omega)_4 \wedge \xi_1
\end{eqnarray*}

where $*(\omega)_4 $ denotes the Hodge dual on $\Omega(M,M_{3}(\mathbb{C}))$. We then find that
\begin{equation}
*(F_{\Omega})=
\left( \begin{array}{cc}
* (F_{\omega})- i(\partial_{\mu} \phi)  *(dx^{\mu} \xi_1) 1_{2 \times 2}&  -(D H)_{\mu}  *(dx^{\mu} \xi_1) \\
(D H)_{\mu}^{\dagger} *(dx^{\mu} \xi_1) & - i(\partial_{\mu} \phi)  *(dx^{\mu} \xi_1)+ * (F_{\alpha}) \end{array}
\right).
\end{equation}

\normalsize
The usual Yang-Mills  type action (\ref{Ya}) involves only four terms
$$S=\dashint \left[(I) + (II) + (III) + (IV)\right] $$
where:
\begin{eqnarray*}
(I)&=& tr(F_{\omega} \wedge *(F_{\omega}))+2 \partial^{\mu} \phi \partial_{\mu} \phi \sqrt{\mid g \mid} d^5x\\
(II)&:=&tr  \left(  (D H)_{\mu}  (D H)^{\mu \dagger}  \right) \sqrt{\mid g \mid} d^5x  \\
(III)&:=&  (D H)_{\mu}^{\dagger} (D H)^{\mu} \sqrt{\mid g \mid} d^5x \\
(IV)&:=& F_{\alpha} \wedge *(F_{\alpha}) +\partial^{\mu} \phi \partial_{\mu} \phi \sqrt{\mid g \mid} d^5x.
\end{eqnarray*}
\normalsize

and $d^{5}x=d^4 x \xi_1$.
Using the fact that $tr(AB^{\dagger})=B^{\dagger}A$ for any $A,B \in M_{2 \times 1}(\mathbb{C})$, we finally find that 
\begin{eqnarray*}
S&=& \dashint\left[ tr(F_{\omega} \wedge *(F_{\omega})) + F_{\alpha} \wedge  *(F_{\alpha}) 
+2 (D H)_{\mu}^{\dagger} (D H)^{\mu}  \sqrt{\mid g \mid} d^5x +3 \partial^{\mu} \phi \partial_{\mu} \phi \sqrt{\mid g \mid} d^5x \right],  
\end{eqnarray*}
\normalsize
i.e., after dimensional reduction,
\begin{equation}
\label{ff}
S=\int_{M}  \left[ tr(F_{\omega} \wedge *(F_{\omega})_{4}) + F_{\alpha} \wedge  *(F_{\alpha})_{4} 
+2 (DH)_{\mu}^{\dagger} (D H)^{\mu}  \sqrt{\mid g \mid} dx^4 + 3 \partial^{\mu} \phi \partial_{\mu} \phi \sqrt{\mid g \mid} d^4x\right].
\end{equation}

The first two terms in (\ref{ff}) are the Yang-Mills actions of the $U(2)$ and $U(1)$ gauge fields.  To recover the classical $SU(2)$ and $U(1)$ gauge field strengths, we further impose the constraint that  
$$tr(\omega)=\alpha.$$

There is no mass term and no quartic potential for the Higgs field, but such terms are gauge-invariant and are generated  under  renormalization. We will elucidate why such terms are absent in section IV.
\bigskip

To determine the Yukawa couplings, we recall the definition of $\tilde{\mathcal{M}}(\mathcal{A})$ and note that $\Omega^{1}(\mathcal{A})_{ V}$ has 5 generators. The Clifford action $c: \Omega^{1}(\mathcal{A})_{ V}  \rightarrow End(\tilde{\mathcal{M}}(\mathcal{A}))$ is then given by 
\begin{eqnarray*}
c(dx^{\mu}):= i\tilde{\gamma}^{\mu}  \otimes 1\\
c(\xi_1 ):=\gamma^{5} \otimes 1
\end{eqnarray*}
where $\tilde{\gamma}^{\mu}$'s are the Dirac matrices in curved spacetime, i.e., $\tilde{\gamma}^{\mu}=e^{\mu}_{a} \gamma^{a}$ with $ \lbrace \gamma^{a}, \gamma^{b} \rbrace=-2 \eta^{ab}$ (we work with the  signature $(-,+,+,+)$),  $\gamma^{a\dagger}=-\gamma^{a}$ for $a=1,2,3$ and $\gamma^{0 \dagger }=\gamma^{0}$; $\gamma^{5}$ is given by the product $\gamma^{5}=i\gamma^{0}\gamma^{1}\gamma^{2}\gamma^{3}$. One checks that $\gamma^{5 \dagger}=\gamma^{5}$, $(\gamma^{5})^{2}=1$. 
The Dirac Operator is given by $$D_c:=c \circ \tilde{\nabla}:  \tilde{\mathcal{M}}(\mathcal{A}) \rightarrow \tilde{\mathcal{M}}(\mathcal{A}).$$ 

This yields
\begin{eqnarray}
D_c&=&i \gamma^{a} e_{a}^{\mu} (\partial_{\mu} - \frac{1}{2}i \omega^{b c}_{\mu} \Sigma_{b c}+ \Omega_{\mu}) +  \gamma^{5}\Omega_{4}  
\end{eqnarray}

where $\Sigma_{ab}=\frac{i}{4} \left[ \gamma_{a}, \gamma_{b} \right]$ and $\omega^{bc}_{\mu}$ are the components of the spin connection.
 We use the notations
\begin{eqnarray*}
\Psi=\left(\begin{array}{c}
\psi_{1L}\\
\psi_{2L}\\
\psi_{3R}
\end{array}
\right), \qquad  &H=\left(\begin{array}{c}
H_{1}\\
H_{2}
\end{array}
\right), \qquad & \bar{\psi}_{i}=\psi_{i}^{\dagger} \gamma^{0}.
\end{eqnarray*}

The fermionic action, defined by 
\begin{equation}
S=\int d^{4}x\sqrt{ \mid g \mid } \bar{ \Psi} D_c \Psi,
\end{equation}

  for arbitrary  $\Psi \in \tilde{\mathcal{M}}(\mathcal{A})$, with   $\bar{\Psi}=(\bar{\psi}_{1L} \bar{\psi}_{2L} \bar{\psi}_{3R})$,  gives rise to the Yukawa  and axion-fermion couplings through the term $ \bar{\Psi} \gamma^{5 } \Omega_{4}\Psi $:
\begin{eqnarray*}
 \bar{\Psi} \gamma^{5 } \Omega_{4}\Psi 
&=& \left(\bar{\psi}_{1L} H_1 \gamma^{5} \psi_{3R} +\bar{\psi}_{2L} H_2 \gamma^{5}\psi_{3R} - \bar{\psi}_{3R} \bar{H}_1 \gamma^{5} {\psi}_{1L}-\bar{\psi}_{3R} \bar{H}_2 \gamma^{5} {\psi}_{2L}\right)+ i  \sum_{i=1}^{3} \bar{\psi}_{i}  \phi  \gamma^{5} \psi_{i} \\
&=& -\left(\bar{\psi}_{1L} H_1 \psi_{3R} +\bar{\psi}_{2L} H_2  \psi_{3R} + \bar{\psi}_{3R} \bar{H}_1 {\psi}_{1L}+\bar{\psi}_{3R} \bar{H}_2 {\psi}_{2L}\right)+ i \sum_{i=1}^{3} \bar{\psi}_{i}  \phi  \gamma^{5} \psi_{i}.\\
\end{eqnarray*}
Coupling constants can be introduced by rescaling the fields.

\section{Remarks and Conclusions}

 We  should  explain why we do not find any quartic and quadratic terms for the Higgs field. For this purpose,  we outline a parallel between our point of view and Connes' point of view of noncommutative geometry. To do so, we introduce the same toy model as Connes did in [\onlinecite{CO1}], (p.563-567). We consider the discrete space $X=\lbrace{a,b\rbrace}$, formed by two separate points. Suppose that there is a complex vector space $W_{a}$ of dimension $n_a$ attached to $a$, and a complex vector space $W_{b}$ of dimension $n_b$ attached to $b$.  $W=W_{a} \oplus W_{b}$ is a  projective finitely generated left $\mathbb{C} \oplus \mathbb{C}$-module, and one could, exactly as  in [\onlinecite{CO1}],   consider the algebra $\mathcal{A}=\mathbb{C} \oplus \mathbb{C}$  and  introduce a connection.  In Connes' formalism, the space of "noncommutative" one forms $\Omega^{1}(\mathcal{A})$ is  2-dimensional. This is the reason why his Yang-Mills action (p.567 of  [\onlinecite{CO1}])  exhibits quartic and quadratic terms that mimic the Higgs potential.\\
  In our approach, we  consider the vector bundle $W$ as a free $\mathbb{C}$-module.  We can choose a basis of sections $a_1,...,a_{n_a},b_{1},...,b_{n_b}$ with $a_{i}  \in W_{a}$ and $b_{j} \in W_{b}$ for all $i,j$. A connection is a linear map that determines the variation of this basis when one moves along space-time. Here it quantifies  the variation due to  jumping from $a$ to $b$. To quantify this jump, we  introduce the one-dimensional vector space $V=\lbrace \lambda \xi_1, \lambda \in \mathbb{C} \rbrace $ spanned by $\xi_1$, where $\xi_1$ plays the role of $dx$  in the direction of the jump.  The variation $\Delta a_i=a'_i-a_i$ can be written $\Delta a_i =\phi_{i1} b_{1}+...+\phi_{in_b} b_{n_b}$, with  $\phi_{ij} \in \mathbb{C}$. In the same way,   $\Delta b_i =\phi'_{i1} a_{1}+...+\phi'_{in_a} a_{n_a}$, with the $\phi'_{ij} \in \mathbb{C}$. The connection  has the form
$$\nabla \left(\begin{array}{c}
a_1\\
..\\
a_{n_a}\\
b_{1}\\
..\\
b_{n_b}
\end{array}\right) = \left(\begin{array}{cccccc}
0&...&0&\phi_{11}&...&\phi_{1n_{b}}\\
...&...&...&...&...&...\\
0&...&0&\phi_{n_{a}1}&...&\phi_{n_{a}n_{b}}\\
\phi'_{11}&...&\phi'_{1n_{a}}&0&..&0\\
...&...&...&...&...&...\\
\phi'_{n_{a}1}&...&\phi'_{n_{b}n_{a}}&0&..&0
\end{array}\right) \xi_1 
\otimes \left(\begin{array}{c}
a_1\\
..\\
a_{n_a}\\
b_{1}\\
..\\
b_{n_b}
\end{array}\right). 
$$

Our approach is thus  different from Connes' approach and leads to different results.  For instance, quartic and quadratic terms in the $\phi_{ij}$'s vanish because $\Omega(\mathcal{B})_{V}$ is one-dimensional. To get such terms one must enlarge $V$, e.g. take  a two-dimensional vector space. We have carried out such generalizations for the Higgs field but they lead to fermion doubling.  Indeed, if $V$ has two generators $\xi_{1}$ and $\xi_{2}$, one can write the connection $\Omega$ of the last section in the form (neglecting the axion field)
\begin{equation}
\Omega=  \left( \begin{array}{cc}
\omega_{  2 \times 2}&0_{ 2 \times 1}\\
0_{ 1 \times 2} & \alpha_{  1 \times1}\\
\end{array}
\right)  \otimes 1 + \left( \begin{array}{cc}
0_{2 \times 2}&H\\
-H^{\dagger}& 0\\
\end{array}
\right) \otimes \xi_1
+ \left( \begin{array}{cc}
0_{2 \times 2}&H'\\
-H'^{\dagger}&0\\
\end{array}
\right) \otimes \xi_2
\end{equation}

where $H'=iH$ if one wants to recover a quartic term for the Higgs field in the action. Then the Clifford action $c:\Omega^{1}(\mathcal{A})_{ V}  \rightarrow End(\tilde{\mathcal{M}}(\mathcal{A}))$ is given by
\begin{eqnarray*}
c(dx^{\mu} ):=\Gamma^{\mu}  \otimes 1\\
c(\xi_1 ):=\Gamma^{5} \otimes 1\\
c(\xi_2 ):=\Gamma^{6} \otimes 1
\end{eqnarray*}

where $\Gamma^{A}$, $A \in \lbrace 0,1,2,3,5,6 \rbrace$, are $8 \times 8$ complex matrices. The number of spinors has to be multiplied by a factor of  two to make sense of $\Gamma^{A} \psi$,  and we end up with fermion doubling.
As there is, a priori, no obstruction against adding gauge-invariant terms to the action, we prefer a five dimensional model.

The introduction of right-handed neutrinos  is possible within our  formalism. The see-saw mechanism ( see [\onlinecite{BO}], [\onlinecite{KI}] for reviews)   furnishes a potential explanation of  the origin of the mass of the left-handed neutrinos of the Standard Model. It is based on the presence, in the action, of  a Majorana mass term for the right-handed neutrinos, of the form $M_{rr} \bar{\nu}_{r} \nu_{r}^{c}$, and a small Dirac mass $m_{lr} \bar{\nu}_{l} \nu_{r} +h.c.$, with $m_{lr} <<M_{rr}$, coming from Yukawa couplings. The mass matrix  can be written in the form
 $$(\bar{\nu}_{l}  \bar{\nu}_{r}^{c}) \left(\begin{array}{cc}
0&m_{lr}\\
m_{lr}^{\dagger}&M_{rr}
\end{array}\right) \left(\begin{array}{c}
\nu_{l}^{c}\\
\nu_{r}
\end{array}\right). $$

The diagonalization of this mass matrix leads to a small mass for the left-handed neutrinos, of the order of  $m_{lr} M_{rr}^{-1} m_{lr}^{\dagger}$, whereas the Majorana masses for the right handed neutrinos are left essentially unchanged. We can introduce a  Dirac mass in our model.  Consider a toy model, where we only add  one right-handed neutrino, described by a Majorana spinor   $\nu_{R}^{c}=\nu_{R}$. On the free $\mathcal{C}^{\infty}(M,\mathbb{C})$-module $\mathcal{C}^{\infty}(M,\mathbb{C}^{2} \oplus \mathbb{C}^{2})$, we can choose the connection
\begin{equation}
\Omega=  \left( \begin{array}{cc}
\omega_{ 2 \times 2}&0_{  2 \times 2}\\
0_{ 2 \times 2} & \left( \begin{array}{cc} \alpha &0\\0&0 \end{array}
\right)\\
\end{array}
\right)  \otimes 1 + \left( \begin{array}{cc}
0_{2 \times 2}&-m_{lr}\\
m_{lr}^{\dagger}& 0\\
\end{array}
\right) \otimes \xi_1.
\end{equation}

This connection leads to a Dirac mass term in the action. 
\begin{acknowledgements}
A. H. C \ is supported in part by the National Science Foundation under Grant
No. Phys-0854779. 
\end{acknowledgements}
\nocite{*}
\bibliographystyle{plain}
\bibliography{axions}

\end{document}